\documentclass[twocolumn]{aastex62}

\newcommand{\jwst}{{\em JWST}}

\graphicspath{{./}{figures/}}

\submitjournal{ApJ Letters}

\shorttitle{PAH features as tracers of star-formation in MIRI-detected galaxies}
\shortauthors{Langeroodi \& Hjorth}
\begin{document}

%\title{PAH features in \jwst\ NIRCam and MIRI broad-band filters as\\tracers of star-formation in intermediate redshift galaxies}
\title{PAH emission from star-forming galaxies in \jwst\ mid-infrared imaging of the lensing cluster SMACS J0723.3$-$7327}
\email{danial.langeroodi@nbi.ku.dk}

\author[0000-0001-5710-8395]{Danial Langeroodi}
\affil{DARK, Niels Bohr Institute, University of Copenhagen, Jagtvej 128, 2200 Copenhagen, Denmark}

\author[0000-0002-4571-2306]{Jens Hjorth}
\affil{DARK, Niels Bohr Institute, University of Copenhagen, Jagtvej 128, 2200 Copenhagen, Denmark}
%\collaboration{(AAS Journals Data Scientists collaboration)}



\begin{abstract}

The mid-infrared spectra of star-forming galaxies (SFGs) are characterized by characteristic broad PAH emission features at 3--20
%3.3, 6.2, 7.7 and 8.6 
$\mu$m. As these features are redshifted, they are predicted to dominate the flux at specific mid-infrared wavelengths, leading to substantial redshift-dependent color variations in broad-band photometry. The advent of \jwst\ for the first time allows the study of this effect for normal SFGs. 
%Here, we present evolutionary tracks in NIRCam and MIRI mid-infrared 4.4, 7.7, 10, 15, and 18 $\mu$m color-color diagrams of SFGs, as a function of redshift. Furthermore, we generate realistic color-color diagrams of magnitude-limited hypothetical surveys by simulating the distribution of SFGs along the evolutionary tracks as implied by the cosmic star-formation history and star-formation rate function. At specific redshifts, SFGs may stand out in the color-color diagrams by several magnitudes. 
Based on spectral energy distribution templates, we here present tracks in mid-infrared (4.4, 7.7, 10, 15, and 18 $\mu$m) color-color diagrams describing the redshift dependence of SFG colors. In addition, we present simulated color-color diagrams by populating these tracks using the cosmic star-formation history and the star-formation rate function.
Depending on redshift, we find that SFGs stand out in the color-color diagrams by several magnitudes. We provide the first observational demonstration of this effect for galaxies detected in the \jwst\ Early Release Observations of the field towards the lensing cluster SMACS J0723.3$-$7327. While the distribution of detected galaxies is consistent with the simulations, the numbers are substantially boosted by lensing effects. The PAH emitter with the highest spectroscopic redshift, detected in all bands, is 
a multiply-imaged galaxy at $z=1.45$. There is also a substantial number of cluster members, which do not exhibit PAH emission, except for one SFG at $z=0.38$. Future wider-field observations will further populate mid-infrared color-color diagrams and provide insight into the evolution of typical SFGs.
\end{abstract}


\section{Introduction} \label{sec:intro}

When excited by radiation in star-forming (SF) regions, polycyclic aromatic hydrocarbons (PAHs) give rise to characteristic emission features in the mid-infrared range, notably at 3.3, 6.2, 7.7, 8.6, and 11.3 $\mu$m. This emission may account for as much as 10--20\% of the infrared luminosity in starburst galaxies, and the 7.7 $\mu$m feature alone may account for 50 \% of the total PAH luminosity \citep{2007ApJ...657..810D,2007ApJ...656..770S,2010ApJ...723..895W,2013ApJ...769...75S}. As such, PAH luminosity is an excellent star-formation rate (SFR) indicator for galaxies in the nearby universe \citep[e.g.,][]{2004A&A...419..501F,2010ApJ...719.1191T,2016ApJ...818...60S,Xie+2019}. 

At low redshifts ($z \lesssim 0.3$), PAH emission based mid-infrared color selections of star-forming galaxies (SFGs) and active galactic nuclei (AGNs) have been carried out using photometry from the {\em Infrared Space Observatory (ISO)} or the {\em Spitzer Space Telescope} \citep{2000A&A...359..887L, 2009ApJS..182..628V}. Due to the limited sensitivity of {\em Spitzer}, observations of PAH features at higher redshifts ($z \sim 0.3-2.8$) have been limited to bright (ultra) luminous infrared galaxies (LIRGs) and AGNs \citep{2008ApJ...675.1171P, 2015ApJ...814....9K}.

The Mid Infrared Instrument (MIRI) aboard \jwst\ offers an unprecedented opportunity to study the properties of much fainter galaxies at mid-infrared wavelengths. Gravitational lensing may provide an additional boost in sensitivity  \citep{2015ApJ...805...79M}. \citet{2017ApJ...849..111K} \citep[see also][]{2018A&A...617A.130I} predicted the properties of galaxies in MIRI bands, with emphasis on high-redshift ($z \sim 1-2$) luminous galaxies and the possible contribution of AGN to their colors \citep[see][for a similar study, focused on {\em Spitzer}/IRAC colors at $z>1$]{2008ApJ...675.1171P}. The combination of depth and photometric coverage provided by \jwst\ broad-band imaging suggest that normal lower-luminousity star-forming galaxies will be detected frequently over a wide redshift range ($z \sim$ 0--2).

Depending on redshift, PAH features will substantially affect the flux of star-forming galaxies in broad MIRI bands. For instance, the 7.7 $\mu$m PAH emission feature is expected to contribute strongly to the flux observed in the MIRI F770W, F1000W, F1500W, or F1800W bands at redshifts 0, 0.3, 1, and 1.3, respectively. Tracing this feature at higher redshifts becomes challenging, as the longer wavelength MIRI bands become less sensitive. Bluer filters (NIRCam F444W or MIRI F560W, F770W) may provide a short-wavelength baseline band which is largely unaffected by PAH emission \citep[although these filters can be sensitive to 3.3 $\mu$m emission, depending on redshift; see, e.g.,][]{2009ApJ...703..270S, 2018A&A...617A.130I}. This unique combination of redder PAH-sensitive and bluer PAH-unaffected filters can be used to identify normal SFGs with PAH-dominated mid-infrared spectra, based on their location in color-color diagrams. 

In this Letter we explore the redshift evolution of the mid-infrared properties of normal SFGs as seen in the broad-band photometry of \jwst. We generate simulated color-color diagrams (in the F444W, F770W, F1000W, F1500W, and F1800W filters) of field SFGs and compare them with those constructed from our analysis of the \jwst\ Early Release Observations \citep{ERO} towards the SMACS J0723.3$-$7327 cluster of galaxies (SMACS J0723 for short) at $z=0.4$, where lensing may enhance the detection of fainter, higher-redshift SFGs. This provides the first observational demonstration of how normal SFGs populate \jwst\ mid-infrared color-color diagrams.

% To do this, we simulate the SFR distribution of SFGs at each redshift, using the empirically-calibrated cosmic star-formation history and the star-formation rate function. The spectral energy distribution (SED) templates of SFGs and the relation between the PAH luminosity and SFR is then used to translate the simulated SFR distribution into magnitude distribution and color-color diagrams. 

% We compare our simulated color-color diagrams with those constructed from the galaxies detected in our analysis of the \jwst\ Early Release Observations towards the SMACS J0723 cluster of galaxies at $z=0.4$ \citep{ERO}. This provides the first observational demonstration of how normal SFGs populate the \jwst\ mid-infrared color-color diagrams.

\section{Tracks in color-color diagrams} \label{sec: tracks}

\begin{figure}
    \centering
    \includegraphics[width=8cm]{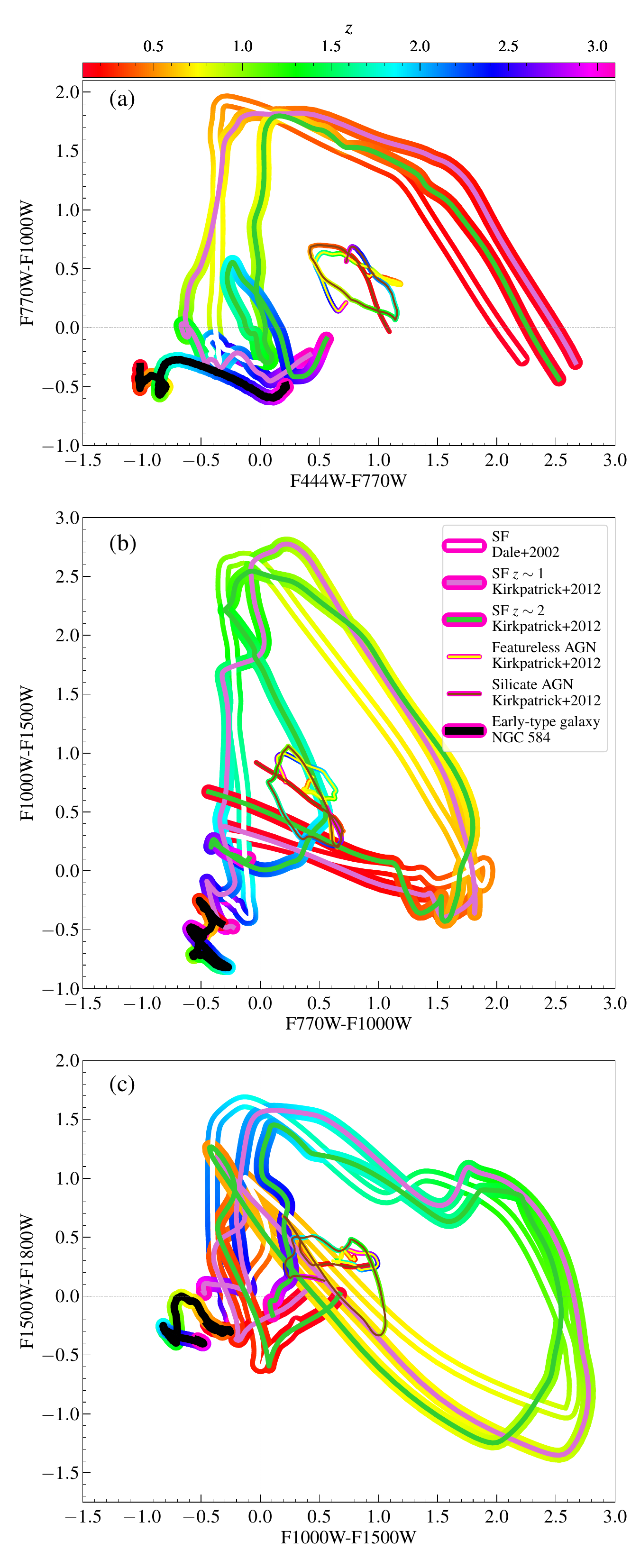}
    \caption{Redshift-dependent tracks of SFGs (thick lines), AGNs (thin lines), and quiescent galaxies (thick lines with thick-black filling) in mid-infrared color-color diagrams.
    From top to bottom: F444W$-$F770W vs.\ F770W$-$F1000W; F770W$-$F1000W vs.\ F1000W$-$F1500W; and F1000W$-$F1500W vs.\ F1500W$-$F1800W color-color diagrams based on SF SED templates from \cite{dale+2002} and \cite{kirkpatrick+2012}; AGN templates from \cite{kirkpatrick+2012}; and the early-type SED of NGC 584.}
    \label{fig: cc unlimited}
\end{figure}

We model the redshift-dependent tracks in the NIRCam and MIRI mid-infrared color-color diagrams using the empirically-calibrated galaxy spectral energy distribution (SED) templates of \cite{dale+2002} and \cite{kirkpatrick+2012}. \cite{dale+2002} created semi-empirical SED templates from $0.36$ to $1100\;\mu$m, constrained in the 3--100 $\mu$m range against IRAS, ISOCAM, and ISOPHOT observations of 69 nearby normal SFGs; in the 52--170 $\mu$m range against ISOLWS data of 228 galaxies; and at $850\;\mu$m against SCUBA observations of 110 galaxies. \cite{dale+2002} parameterized their semi-empirical spectra based on the far-infrared flux density ratio $f_{v} (60\mu$m$)/f_{v} (100\mu$m$)$ (ranging from 0.3 to 1.6; providing an indicator of dust grain temperature), generating templates that describe the observed spectra of normal SFGs spanning a wide range in far-infrared colors and dust temperature. 

\cite{kirkpatrick+2012} created composite SED templates in the rest-frame 0.3--600 $\mu$m range for four sub-samples of $24\;\mu$m-selected $z \sim 0.5$--4 (U)LIRGs in the GOODS-N and ECDFS fields, classified based on a mid-infrared spectral decomposition. These sub-samples consist of $z \sim 1$ SF-dominated galaxies (SF $z \sim 1$, for short), $z \sim 2$ SF-dominated galaxies (SF $z \sim 2$), AGN-dominated galaxies with clear $9.7\;\mu $m silicate absorption (silicate AGN), and AGN-dominated galaxies with featureless mid-infrared spectra (featureless AGN). The SED template calibration sample consists of 95 galaxies with complete spectral coverage from IRAC, IRS, and MIPS on {\em Spitzer}, {\em Herschel} PACS and SPIRE, and ground-based $870\;\mu$m and $1.15$ mm.

We calculate the \jwst\ mid-infrared multi-band magnitudes for an SED template at a given redshift, using the \texttt{SEDPY} package \citep{2019ascl.soft05026J, 2021ApJS..254...22J} to convert the redshifted spectra into response curve-determined mean flux in each band. Motivated by the SMACS J0723 observations (see Section \ref{sec: smacs}), we generate F444W$-$F770W vs.\ F770W$-$F1000W; F770W$-$F1000W vs.\ F1000W$-$ F1500W; and F1000W$-$F1500W vs.\ F1500W$-$F1800W color-color diagrams. These combinations are chosen such that each maximizes the color variations at a particular redshift range by including a relatively blue baseline PAH-unaffected filter as well as two redder filters tracing different PAH features. The F1800W filter allows tracing the strong 6.2 and $7.7 \mu$m PAH features up to $z \sim 1.7$ and $z \sim 1.3$, respectively.

%We exclude the combinations containing MIRI F560W, because real deep multi-band \jwst\ data containing this filter is not yet publicly available (see Section \ref{sec: smacs}). 
%Since the sensitivity of MIRI filters decrease with increasing wavelength, we do not present color-color diagrams consisting of filters redder than 18 $\mu$m.

% \footnote{these color-color diagrams can be found in both PDF and machine-readable formats at {\tt here}}

To generate the color-color diagrams we sample the redshift space from $z = 0.1$ to $z = 3.1$ in 0.01 intervals. At each sampled redshift, we calculate the colors for each of the SF-dominated SED templates in \cite{kirkpatrick+2012} as well as the SF template from \cite{dale+2002} with $f_{\nu} (60\mu \textrm{m})/f_{\nu} (100\mu \textrm{m}) = -0.5232$. For each SF template, Figure \ref{fig: cc unlimited} presents the color-color space tracks in the $z =$ 0.1--3.1 range. In each diagram, the location of the SFGs depend strongly on their redshifts. At higher redshifts they settle in a stationary locus at the lower left of the diagrams, at slightly negative colors.
%stand out at a specific redshift range - where their location on the color-color diagram is highly sensitive to redshift - before settling into a stationary locus to the bottom-left of the (0,0) point.

We also generate $z = 0.1$--3.1 tracks for AGNs and quiescent galaxies to compare with those of SFGs (Figure \ref{fig: cc unlimited}). AGNs are represented by the AGN-dominated SED templates of \cite{kirkpatrick+2012}, while quiescent galaxies are represented by the SED of the early-type galaxy NGC 584 \citep{2014ApJS..212...18B}. AGNs populate a region in the diagrams that is generally distinct from SFGs and quiescent galaxies at all redshifts. Quiescent galaxies populate the region toward the bottom-left of the stationary locus of the SF tracks. Their location are roughly independent of redshift because of their approximate power-law spectral energy distributions.

\section{Simulated color-color diagrams} \label{sec: simulations}

We next simulate realistic color-color diagrams of SFGs\footnote{AGNs are not included in these simulations because only a few AGNs are expected to be detected per MIRI FOV. See Section \ref{sec: discussion and conclusion} for a more detailed discussion.} by populating the redshift-dependent tracks taking into account the cosmic star-formation history, the SFR of individual galaxies, and the sensitivity of the \jwst\ instruments. As described below, we first simulate the SFR distribution of individual galaxies from $z = 0.1$ to $z = 3.1$ over a given field of view (FOV). Then we convert the simulated redshift and SFR distribution into mid-infrared multi-band magnitudes, using the SF SED templates normalized according to the correlation between the SFR and PAH luminousity of SFGs. 

Simulating the redshift and SFR distribution of individual galaxies requires knowledge of the total SFR at each redshift interval, as well as the distribution of star-formation over individual galaxies at that redshift. We sample the redshift space in 0.01 intervals, using the cosmic star-formation history relation of \cite{Madau+2014} to calculate the total SFR of the FOV at each redshift bin. In each bin, we then randomly distribute the total SFR over individual galaxies, using the interpolated $\textnormal{SFRF}(z)$ (star formation rate function) from \cite{Sobral+2014} ($z < 2.23$) and \cite{Smit+2012} ($z > 2.23$). While \cite{Madau+2014} and \cite{Smit+2012} adopt a Salpeter IMF, \cite{Sobral+2014} use a Chabrier IMF. We scale from the former to the latter by multiplying the SFRs by a factor of 1.59. 

% We assign the highest redshift of each bin as the redshift of the drawn SFRs.  
% Although \cite{Sobral+2014} calibrated the $\textnormal{SFRF}(z)$ at $z = $ 0.4, 0.84, 1.47, and 2.23, its interpolation over redshift is shown to be consistent with both the lower and higher redshift results.

We convert the drawn redshift and SFR distribution into mid-infrared intrinsic luminosity of SFGs, adopting a linear correlation between the SFR and PAH luminosity ($L_{\textrm{\scriptsize PAH}}$) as calibrated in \cite{Xie+2019}. \cite{Xie+2019} divided their sample based on stellar mass ($M_{\star}$) and showed that galaxies with $M_{\star} > 10^{9}M_{\odot}$ are in general 100 times brighter in $L_{\textrm{\scriptsize PAH}}$ than $M_{\star} < 10^{9}M_{\odot}$ galaxies with the same SFR (see their Figure 3 and Table 3). To make this distinction, we estimate the stellar mass corresponding to the drawn SFRs, using the star-formation main sequence (MS) calibration from \cite{2014ApJS..214...15S}. For each drawn redshift and SFR, we randomly draw the stellar mass in a $1\sigma$ truncated 0.2 dex normal distribution centred on the mean stellar mass determined by the best-fit MS relation. The dispersion is motivated by the true scatter of the MS relation. Since \cite{2014ApJS..214...15S} adopt a Kroupa IMF, we multiply their stellar mass by 1.55 to convert from the Kroupa IMF to the Salpeter IMF. The resulting sample of drawn SFR, redshift, and stellar mass is then converted into PAH 7.7 $\mu$m luminosity, using the \cite{Xie+2019} conversion. 

For each drawn SFR and redshift, we randomly select a SF SED template (see Section \ref{sec: tracks}), re-scale the template to match the drawn intrinsic ${\textnormal{PAH}\;7.7\ \mu \textrm{m}}$ luminosity, and calculate the \jwst\ mid-infrared magnitudes for the redshifted spectra following the method of Section \ref{sec: tracks}. 
% We generate sample surveys with smaller FOVs and limited depth by randomly selecting a sub-sample from the simulated 100 arcmin$^2$ FOV. 
The top panels in Figures \ref{fig: smacs 444}, \ref{fig: smacs 770}, and \ref{fig: smacs 1000} show the resulting color-color diagrams for a MIRI FOV, consisting only of SFGs with observed magnitudes below the limiting magnitudes of the \jwst\ observations of the SMACS J0723 field (see Section \ref{sec: smacs}). We expect 18, 15, and 11 SFGs per MIRI FOV to enter the diagrams in Figures  \ref{fig: smacs 444}, \ref{fig: smacs 770}, and \ref{fig: smacs 1000}, respectively.
% In Figures \ref{fig: smacs 444}, \ref{fig: smacs 770}, and \ref{fig: smacs 1000}, we expect 790 (18), 640 (15), and 490 (11) SFGs to be detected simultaneously in all three bands per 100 arcmin$^2$ (MIRI FOV).

% For each drawn SFR and redshift, we randomly select a SF SED template (see Section \ref{sec: tracks}), re-scale the template to match the drawn intrinsic ${\textnormal{PAH}\;7.7\ \mu \textrm{m}}$ luminosity, and calculate the \jwst\ mid-infrared magnitudes for the redshifted spectra following the method of Section \ref{sec: tracks}. We generate sample surveys with smaller FOVs and limited depth by randomly selecting a sub-sample from the simulated 100 arcmin$^2$ FOV. The top panels in Figures \ref{fig: smacs 444}, \ref{fig: smacs 770}, and \ref{fig: smacs 1000} show the resulting color-color diagrams for a MIRI FOV, consisting only of SFGs with observed magnitudes below the limiting magnitudes of the \jwst\ observations of the SMACS J0723 field (see Section \ref{sec: smacs}). In Figures \ref{fig: smacs 444}, \ref{fig: smacs 770}, and \ref{fig: smacs 1000}, we expect 790 (18), 640 (15), and 490 (11) SFGs to be detected simultaneously in all three bands per 100 arcmin$^2$ (MIRI FOV).

\section{color-color diagrams for the\\SMACS J0723.3$-$7327 field} \label{sec: smacs}

As a proof of concept, we analyze the NIRCam and MIRI Early Release Observations of the SMACS J0723 field \citep[PID 2736, PI: Pontoppidan;][]{ERO}. The broad-band imaging consists of 6 overlapping NIRCam images in the F090W, F150W, F200W, F277W, F356W, and F444W filters reaching a point source sensitivity of $\textnormal{AB}\sim29.8$ mag, and 4 overlapping MIRI images in the F770W, F1000W, F1500W, and F1800W filters reaching $5\sigma$ depths of $\textnormal{AB}\sim 26.3$, 25.1, 24.0, and 23.0 mag, respectively. For each filter, we acquire the photometry- and astrometry-calibrated data products from the Mikulski Archive for Space Telescopes (MAST), reduced using the 1.5.3 version of the \jwst\ Calibration Pipeline\footnote{\url{https://jwst-pipeline.readthedocs.io}}. Here, we focus on the imaging in the mid-infrared bands F444W, F770W, F1000W, F1500W, and F1800W.

\begin{figure}
    \centering
    \includegraphics[width=8.5cm]{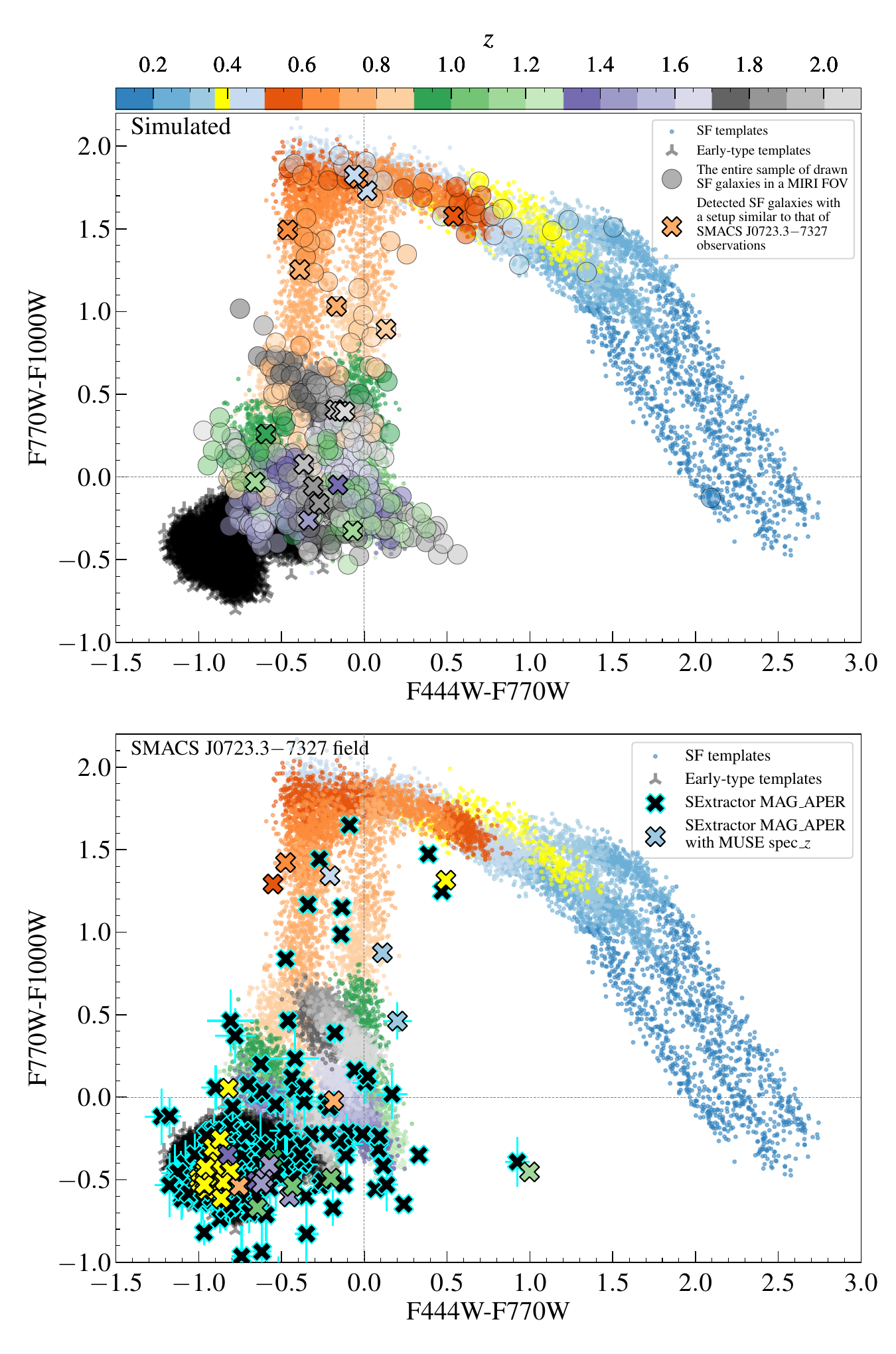}
    \caption{F444W$-$F770W vs.\ F770W$-$F1000W color-color diagram for the \jwst\ observations of the SMACS J0723 field. 
    Top panel: Simulated diagram for SFGs detected (filled crosses) in a MIRI FOV with the same magnitude limits as those of the SMACS J0723 observations. The open circles indicate the entire sample of SFGs in this FOV, most of which are undetected.
    %due to low SFR/PAH luminosity. 
    The color-shaded tracks (small filled circles) are similar to the tracks of Figure \ref{fig: cc unlimited}, with an additional 0.05 mag scatter. The shaded black region shows the expected location of early-type galaxies. We expect to detect 18 SFGs per MIRI FOV. 
    Bottom panel: Color-color diagram of 173 galaxies detected in \jwst\ imaging (filled crosses), supplemented with MUSE spectroscopic redshifts where available (color-filled crosses). We identify 23 cluster galaxies with available spectroscopic redshifts ($z = 0.387\pm{0.020}$; yellow-filled crosses). There is only one confirmed cluster galaxy that resembles the color-color diagram location of SFGs (yellow-filled cross at (0.5,1.3); see Section \ref{sec: discussion and conclusion}).}
    \label{fig: smacs 444}
\end{figure}

There is a significant astrometric offset (more than $1\arcsec$) between the NIRCam imaging and the MIRI imaging data products available on MAST. We correct this by re-aligning the image in each of the F356W\footnote{The F356W imaging is only used to search for AGN candidates (see Section \ref{sec: discussion and conclusion}), and not in the production of color-color diagrams.}, F444W, F770W, F1000W, F1500W, and F1800W filters directly to {\em Gaia} stars, using the \texttt{GAIA Starlink} software. In the resulting re-calibrated images, the centroids of Source Extractor-identified \citep[\texttt{SExtractor;}][]{SEX} point sources overlap within $0.1\arcsec$ across all the filters. We further calibrate the astrometry by correcting the sky position of \texttt{SExtractor}-detected stars (\texttt{CLASS$\_$STAR} $> 0.5$) in F356W and F444W imaging against their position in MIRI imaging 
to achieve a sub-MIRI pixel offset between the point source centroids in the F356W/F444W and MIRI imaging. This astrometric accuracy is sufficient for the purpose of cross-matching the detections in F444W imaging with those in MIRI imaging.

\begin{figure}
    \centering
    \includegraphics[width=8.5cm]{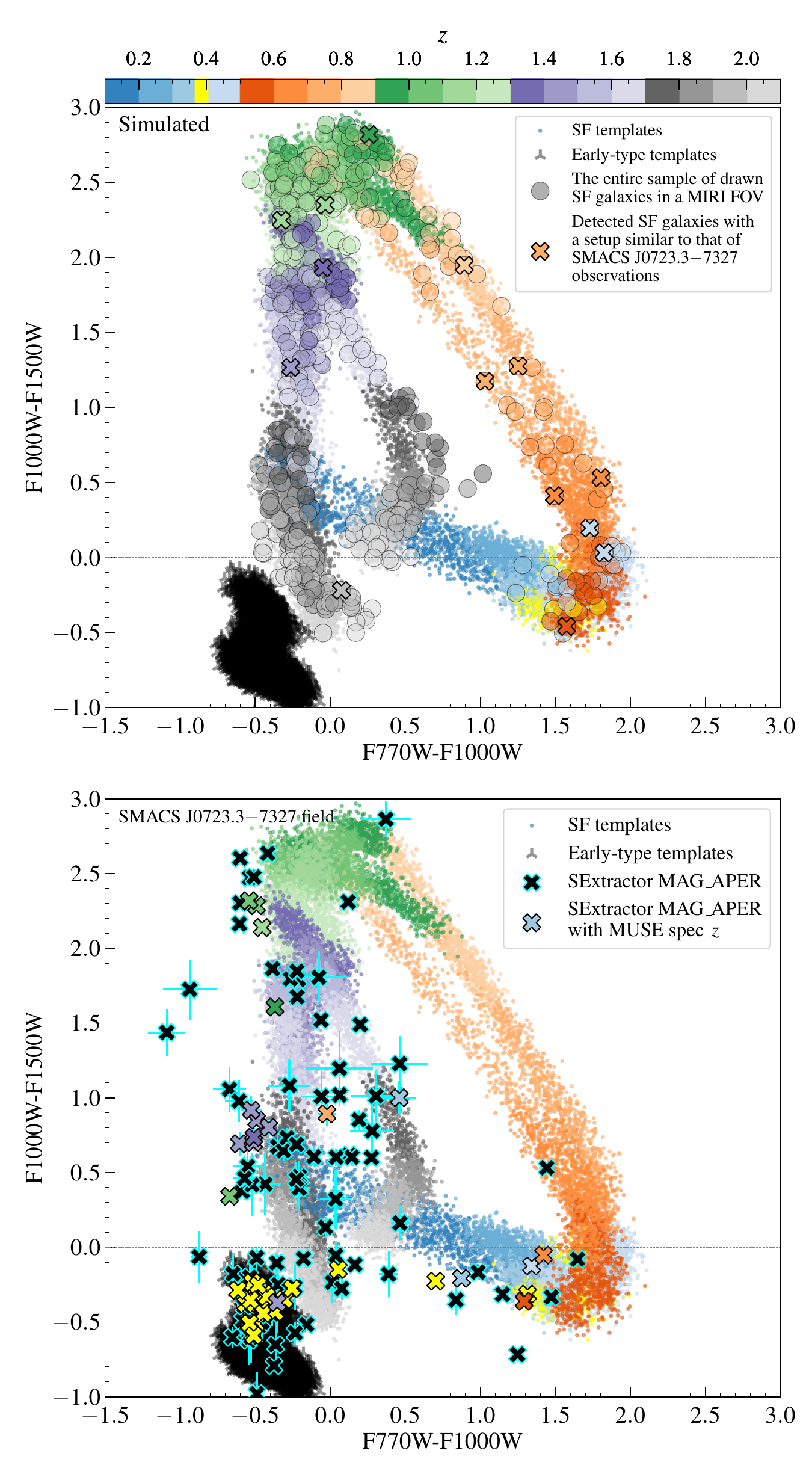}
    \caption{Same as Figure \ref{fig: smacs 444}, but showing the F770W$-$F1000W vs.\ F1000W$-$F1500W  colors for a total of 129 detections.}
    \label{fig: smacs 770}
\end{figure}

We identify the bright sources in MIRI filters by running \texttt{SExtractor} in double-image mode. We use the imaging in F770W as the detection image and adopt a \texttt{SExtractor} setup corresponding to a signal-to-noise ratio (SNR) of 2, setting \texttt{DETECT$\_$MINAREA} = 4 and \texttt{DETECT$\_$THRESH} = 1. This results in forced photometry of F770W detections over the same sky segment across all MIRI filters. Similarly, we measure the photometry of bright F356W and F444W sources by running \texttt{SExtractor} in double-image mode with the F444W filter as the detection image, \texttt{DETECT$\_$MINAREA} = 4 and \texttt{DETECT$\_$THRESH} = 1 (SNR = 2). We cross-match the NIRCam and MIRI detections by matching the \texttt{SExtractor}-determined source centroid sky positions of the NIRCam detections against MIRI detections within a $0.3\arcsec$ radius. 

\begin{figure}
    \centering
    \includegraphics[width=8.5cm]{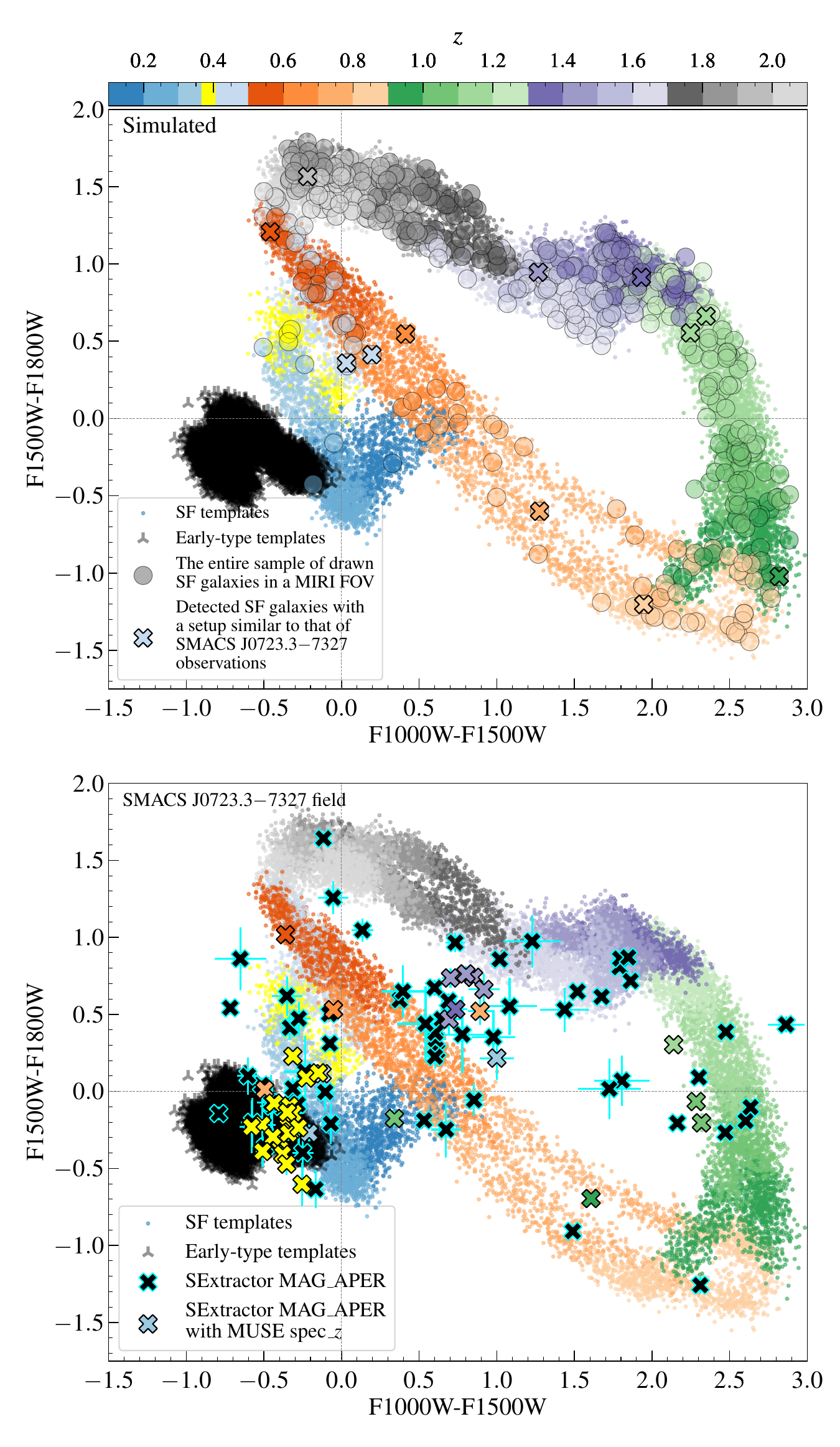}
    \caption{Same as Figures \ref{fig: smacs 444} and \ref{fig: smacs 770}, but showing the F1000W$-$F1500W vs.\ F1500W$-$F1800W colors for a total of 99 detections.}
    \label{fig: smacs 1000}
\end{figure}

% We have made the assumption that the galaxy colors are spatially uniform, which implies that the measured colors should be robust to the choice of aperture as long as the same aperture is used in every filter and the chosen aperture is sufficiently large to avoid saturation limits.
We use the \texttt{SExtractor}-measured aperture photometry (\texttt{MAG$\_$APER}) within a $1\arcsec$ diameter as the measured magnitude in each filter. To check the robustness of colors measured in this way, we compare the \texttt{MAG$\_$APER} colors with \texttt{SExtractor}-measured \texttt{MAG$\_$AUTO} colors, which are measured by adopting a flexible aperture size determined for the source image in each filter by the second order moments of its light distribution. Generating the color-color diagrams with \texttt{MAG$\_$AUTO} colors instead of \texttt{MAG$\_$APER} colors has no significant effect on the distribution of galaxies, and only slightly increases their scatter with respect to the simulated redshift-dependent tracks. The bias between the \texttt{MAG$\_$AUTO} and \texttt{MAG$\_$APER} colors is small ($ < 0.05$ mag for all colors).

We use the most recent NIRCam flux calibration (\texttt{jwst$\_$0989.pmap}) to determine the zero-point magnitudes of F356W and F444W imaging. 
The photometry of MIRI imaging was calibrated using the default \texttt{jwst$\_$0916.pmap} MIRI flux calibration used in the 1.5.3 version of the \jwst\ calibration pipeline; this calibration is the same as the most recent calibration, \texttt{jwst$\_$0989.pmap}.
The mid-infrared color-color diagrams of the detected galaxies (filled crosses) are presented in the bottom panels of Figures \ref{fig: smacs 444}, \ref{fig: smacs 770}, and \ref{fig: smacs 1000}. In each diagram, we only plot the galaxies (\texttt{CLASS$\_$STAR} $< 0.5$) where the estimated error of aperture photometry (\texttt{MAGERR$\_$APER}) is below 0.2 mag (i.e., a $\sim 5\sigma$ detection) in all three bands.

Where available, we assign spectroscopic redshifts to the detected galaxies. The SMACS J0723 field was observed under the MUSE 0102.A-0718 program \citep{caminha+2022}, resulting in a spectroscopic redshift catalog containing 78 secure redshift measurements of {\em HST}-detected \citep[Reionization Lensing Cluster Survey program,][]{2019ApJ...884...85C} sources. We re-calibrate the astrometry of this catalog against our calibration of the NIRCam and MIRI imaging by correcting the sky position of MUSE stars against that of F444W stars (F356W and F444W contain the highest number of star detections, resulting in 7 matches with the MUSE catalogue). We cross-match the MUSE catalog with the galaxies detected in F444W imaging, within a $0.3\arcsec$ radius. This results in assigning spectroscopic redshifts to 42, 41, and 36 galaxies in the diagrams of Figures \ref{fig: smacs 444}, \ref{fig: smacs 770}, and \ref{fig: smacs 1000}, respectively (color-filled crosses). Additionally, \cite{2022arXiv220708778C} measured 10 new spectroscopic redshifts for galaxies in the SMACS J0723 field based on the \jwst\ NIRSpec medium-resolution data (PID: 2736). One of these galaxies at $z = 1.163$ \citep[ID: 9483 in][]{2022arXiv220708778C} enters the SMACS J0723 color-color diagrams (brown-filled cross).

\section{Discussion and Conclusion} \label{sec: discussion and conclusion}

\begin{figure*}
    \centering
    \includegraphics[width=18cm]{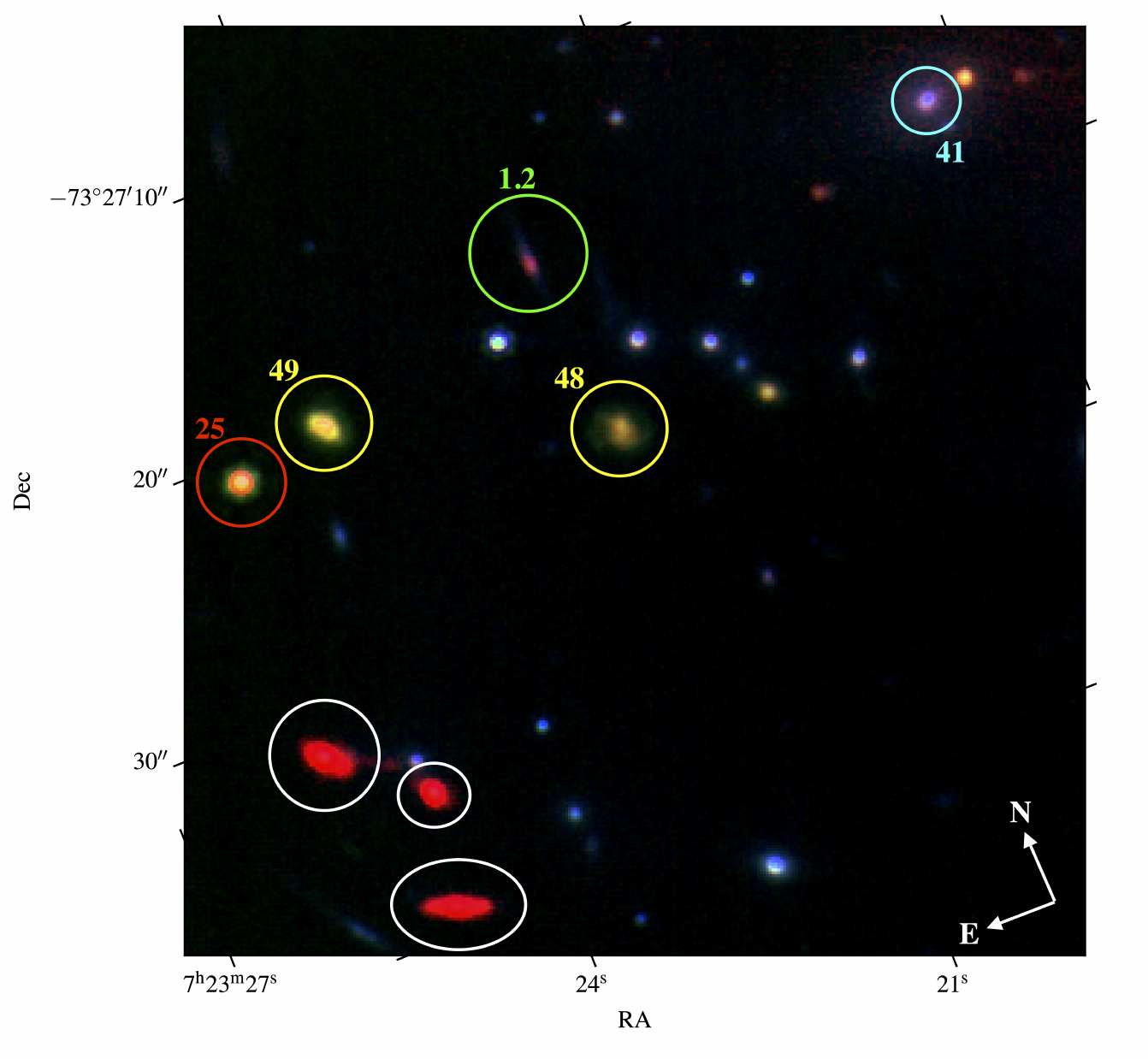}
    \caption{Color-composite image (F1500W: red, F1000W: green, F770W: blue) of a $320\times310$ pixel ($0.11\arcsec$ per pixel) except of the \jwst/MIRI SMACS J0723 observations \citep[see][for the full FOV]{ERO}. The number next to each circle indicates the ID in the spectroscopic redshift catalog of \cite{caminha+2022}. 
    The yellow circles mark two SFGs at $z =$ 0.4233 (ID: 48) and 0.5189 (ID: 49). The light-blue circle marks the brightest cluster (elliptical) galaxy (ID: 41). The green circle shows a lensed source at $z = 1.4503$ (ID: 1.2). We identify one cluster ($z = 0.3822$; ID: 25) SFG which is marked with the red circle.
    The white circles mark three SFGs at estimated $z \sim 0.9$--1.3, as indicated by their location in the color-color diagrams.}
    \label{fig: rgb}
\end{figure*}

Overall, the detected galaxies occupy the regions of the color-color diagrams as predicted: SFGs in a specific redshift range stand out by several magnitudes from the bulk of detections occupying the region at the bottom left of the flat-spectrum $(0,0)$ point, presumably populated by quiescent galaxies and higher redshift SFGs. Three non-cluster galaxies with available MUSE spectroscopic redshifts ($z = $ 0.42, 0.52, 0.60; IDs\footnote{Unless otherwise specified, throughout this work ID refers to ID in the spectroscopic redshift catalogue of \cite{caminha+2022}.}: 48, 49, 50) stand out by $\sim 1.5$ mag at the upper region of the F444W$-$F770W vs.\ F770W$-$F1000W color-color diagram (bottom panel of Figure \ref{fig: smacs 444}). We confirm their classification as SFGs with spiral morphology through visual inspection. Two of these galaxies are marked with yellow circles in Figure \ref{fig: rgb}. 

For the cases where spectroscopic redshift is not available, the location of SFGs in the color-color diagrams can strongly constrain their redshifts. In Section \ref{sec: simulations} (see also Figure \ref{fig: cc unlimited}), we showed that the MIRI colors of SFGs can be highly sensitive to their redshifts. Therefore, photometric redshifts of SFGs can be estimated by comparing their location on the MIRI color-color diagrams with the synthetic redshift-dependent tracks. For instance, the white circles in Figure \ref{fig: rgb} show three SFGs with estimated redshifts of $z \sim 0.9$--1.3. In Langeroodi et al. (2023, in prep.)\ we assess the accuracy of such mid-infrared derived photometric redshifts.

In Figure \ref{fig: smacs 444}, we identify 23 galaxies associated with the cluster (yellow-filled crosses) based on MUSE spectroscopic redshifts ($z = 0.387 \pm 0.02$); this accounts for all the cluster galaxies identified in \cite{caminha+2022}. In Figures \ref{fig: smacs 770} and \ref{fig: smacs 1000}, we identify 23 and 19 cluster galaxies, respectively. Almost all of the cluster galaxies (except for one; see below) occupy the region corresponding to quiescent galaxies, which is expected due to the higher efficiency of various quenching channels in cluster environments \citep[see, e.g.,][]{2010MNRAS.404.1231V, 2017MNRAS.470.4186B, 2021MNRAS.506.4760D, 2022arXiv220712491K, 2022A&ARv..30....3B}. The light-blue circle in Figure \ref{fig: rgb} marks the brightest cluster galaxy (ID: 41). The excellent agreement between the color-color diagram region occupied by cluster early-type galaxies and their predicted location (see Section \ref{sec: tracks} and Figure \ref{fig: cc unlimited}), as is especially evident in Figures \ref{fig: smacs 770} and \ref{fig: smacs 1000}, verifies the accuracy of the adopted relative flux calibration for the MIRI filters (see Section \ref{sec: smacs}).

We detect two galaxies located in front of the cluster (dark-orange crosses in each color-color diagram; ID: 21 and 22), which are consistent neither with the predicted colors of SFGs at their corresponding spectroscopic redshifts nor with early-type galaxies. We interpret these as SFGs with a significant contribution of quenched SED to their spectra. 
The linear combination of a quenched spectrum with that of a SFG results in moving the color-color diagram location of the galaxy toward the stationary locus of the quenched galaxies. Therefore, rather than being exclusively confined to the presented tracks, some galaxies with significant quenched fraction end up populating the `forbidden' region in-between. 

%Since the early-type spectra can in general be modelled with a negative-sloped power law in the 1--30 $\mu$m range, its contribution to the PAH-dominated spectra of SFGs in a similar frequency range results in partial smoothening of the PAH-features \citep[see e.g. ][]{2014ApJS..212...18B}. This can potentially result in moving the galaxy toward the location of early-types in the color-color diagram. 

The number counts of detected SFGs are in general agreement with our simulations. We expect fewer than one $z < 0.3$ and a handful of $0.3 < z < 0.5$ SFGs to enter the color-color diagrams, both of which are confirmed by the SMACS J0723 diagrams. However, we detect more galaxies than predicted at $z > 0.5$, where lensing starts to influence the number counts. We suspect that the excess detections are primarily caused by a combination of lensing magnification and multiple imaging \citep{2015ApJ...805...79M}. Two spectroscopically-confirmed multiply-imaged galaxies enter the diagram in Figure \ref{fig: smacs 444} including the $z = $ 1.4503 1.1, 1.2, and 1.3 images from \cite{caminha+2022} as well as the $z = $ 1.3782 2.1 and 2.3 images. The 1.1, 1.2, and 1.3 images also enter the diagrams in Figures \ref{fig: smacs 770} and \ref{fig: smacs 1000}. The 1.2 lensed image is marked with the green circle in Figure \ref{fig: rgb}. Early-type galaxies with non-negligible star-formation can contaminate the $z > 0.5$ SFG number counts as well, especially in the F444W$-$F770W vs.\ F770W$-$F1000W diagram where the early-type region is close to the $z > 0.5$ SFG region. 

We note that the method used herein for simulating the detected population of SFGs (see Section \ref{sec: simulations}) relies on the SFR--$L_{\textrm{\scriptsize PAH}}$ calibration of \cite{Xie+2019}, which is derived from a sample of galaxies in the nearby Universe ($z < 0.3$). In particular, \cite{Xie+2019} suggested that at a fixed SFR, galaxies with $M_{\star} < 10^9 M_{\odot}$ are on average $\sim 100$ times fainter in $L_{\textrm{\scriptsize PAH}}$ compared to galaxies with $M_{\star} > 10^9 M_{\odot}$; we assume that this is valid at redshifts beyond $z = 0.3$. This assumption strongly constrains the simulated number of SFG detections at $z > 0.5$, where at a fixed SFR the stellar mass of galaxies are on average at least 1.6 dex lower than at $z = 0$ \citep[see][]{2014ApJS..214...15S}. The extrapolation of this $10^9 M_{\odot}$ threshold to redshifts higher than $z = 0.3$ should be further investigated in field observations, where unlike the SMACS J0723 field the number counts are not boosted by lensing or dominated by quenched cluster galaxies.

Variations in the relative strengths of PAH emission features can potentially increase the observational scatter of SFGs with respect to the simulated redshift-dependent tracks in mid-infrared color-color diagrams. The relative strengths of PAH emission features are a function of the ionization state of grains and their size distribution \citep{2001ApJ...551..807D, 2020MNRAS.494..642M}; this is especially pronounced for the $3.3 \mu$m and $11.3 \mu$m features. Since the $3.3 \mu$m feature at most contributes $\sim 0.07$ mag to the measured F444W or MIRI magnitudes (depending on the filter and redshift), variations in the relative strength of this feature do not catastrophically propagate into galaxy colors. In contrast, the $11.3 \mu$m feature can contribute as much as $\sim 0.3$ mag to the measured MIRI magnitudes; the pessimistic theoretically suggested 1 dex range for the relative strength of this feature \citep[as shown in][]{2020MNRAS.494..642M} can result in a $\sim 0.2$ mag scatter in mid-infrared colors. However, this entire range is not populated by observed SFGs.

Compared to the theoretically suggested range, observed SFGs in the nearby Universe ($z \sim 0$) exhibit a much smaller variation in the relative strengths of PAH emission features \citep{2007ApJ...656..770S, 2009ApJ...705..885O, 2010ApJ...723..895W, 2014ApJ...790..124S}. For instance, in the \cite{Xie+2019} sample at $z < 0.3$ the line ratio of the $11.3 \mu$m to $7.7 \mu$m features varies by only $\sim 0.1$ dex ($1\sigma$ distribution), which translates into a negligible 0.04 mag scatter in the observed MIRI colors of SFGs (smaller than the scatter implemented in the simulated tracks of Figures \ref{fig: smacs 444}, \ref{fig: smacs 770}, and \ref{fig: smacs 1000}). 
There is evidence that the scatter in the relative strengths of PAH features is larger at $z > 1$ compared to the nearby Universe \citep{2020ApJ...892..119M}. \cite{2020ApJ...892..119M} measured a 0.2 dex scatter in the line ratio of the $11.3 \mu$m to $6.2 \mu$m features at $z > 1$, which can translate into a 0.1 mag scatter in the measured MIRI colors. This might be affecting the location of SFGs in the color-color-diagrams of Figures \ref{fig: smacs 770} and \ref{fig: smacs 1000}, as is evident especially at $z > 1$ by the larger scatter with respect to the simulated tracks.

% Moreover, there is evidence that the scatter in the relative strengths of PAH features is larger at $z > 1$ \citep{2020ApJ...892..119M} compared to $z \sim 0$ \citep{2007ApJ...656..770S, 2009ApJ...705..885O, 2010ApJ...723..895W, 2014ApJ...790..124S}. The measured 0.2 dex scatter \citep{2020ApJ...892..119M} of the $11.3 \mu$m to $6.2 \mu$m feature line ratio at $z > 1$ can translate into a 0.1 mag scatter in the measured colors. This might be affecting the location of SFGs in the color-color-diagrams of Figures \ref{fig: smacs 770} and \ref{fig: smacs 1000}, as is evident especially at $z > 1$ by the larger scatter with respect to the template tracks.
% The contribution is particularly strong for SFGs at redshifts where the $11.3 \mu$m feature is traced by one of the MIRI filters; i.e. at $z \sim 0.2$ -- 0.5 and $z \sim 0.5$ -- 0.8 in the F1500W and F1800W filters, respectively.

\cite{2022arXiv220906219L} predicted a few ($\sim 4$) AGN detections per MIRI FOV. 
We detect one galaxy with strong point source emission in MIRI imaging ($\nu L_\nu \approx 4\times10^{42}$ erg s$^{-1}$ at $10^{14}$ Hz). It has a spectroscopic redshift consistent with it being a cluster member ($z = 0.3822$; ID: 25; marked with the red circle in Figure \ref{fig: rgb}). The location of this galaxy in the color-color diagrams is consistent with that of an SFG at its redshift. Therefore, from a color-color diagram point of view, this source could be an AGN in a star-forming galaxy. Such systems have recently been shown to resemble SFG PAH emissions \citep{2022arXiv220811620G, 2022arXiv220906741L, 2022arXiv221004647D}, if the effect of AGN radiation fields on the excitation of PAHs in nuclear star-forming regions are not particularly strong \citep[e.g.][]{2022arXiv220906741L}. However, we do not find any X-ray detection associated with this source, and its F356W$-$F444W NIRCam color ($-0.21 \textnormal{ mag } \pm 0.01$) is not consistent with the W1$-$W2 $>0.8$ AGN selection criterion \citep[see, e.g.,][]{2012ApJ...753...30S}. In the entire sample shown in Figure \ref{fig: smacs 444} and Table \ref{table: photometry}, we do not find any source with F356W$-$F444W$>0.8$.
%However, we do not detect any galaxy as a strong AGN candidate purely based on its location on the color-color diagrams. For further confirmation, we also studied the color-color diagram of the sources that were excluded by \texttt{SExtractor} as stars, the color-color diagram location of none of which resembles that of AGNs.

This Letter has presented the first demonstration of \jwst\ era color-selection of SFGs based on their PAH emission. Upcoming deeper/wider \jwst\ mid-infrared imaging will further populate the color-color diagrams presented herein, and allow the construction of large samples of normal SFGs (as well as quenched galaxies) at redshifts previously unexplored. Such samples can be instrumental in population studies of normal SFGs.

\begin{acknowledgments}
The authors would like to thank the anonymous referee who greatly helped in improving the presentation and discussion. The authors would like to acknowledge initial discussions with Luis Colina and Hiroyuki Hirashita, as well as constructive comments from Sandra Raimundo. This work was supported by a VILLUM FONDEN Investigator grant (project number 16599). The \jwst\ data used in this work were obtained from the Mikulski Archive for Space Telescopes (MAST) at the Space Telescope Science Institute. The specific NIRCam and MIRI observations analyzed can be accessed via\dataset[10.17909/v0ka-bd26]{https://doi.org/10.17909/v0ka-bd26}. 
\end{acknowledgments}

\bibliography{apjl}
\bibliographystyle{aasjournal}

\startlongtable
\begin{table*}
\centering
\begin{tabular}{ |l|l|l|l|l|l|l|l|l| } 
 \hline
ID & RA & DEC & F444W & F770W & F1000W & F1500W & F1800W & spec-$z$ \\ 
\hline\hline
130 & $110.8498258$ & $-73.4625933$ & $24.94 \pm 0.01$ & $24.74 \pm 0.09$ & $24.28 \pm 0.07$ & $23.28 \pm 0.09$ & $23.07 \pm 0.11$ & 0.3217\\ 
158 & $110.8488494$ & $-73.4607803$ & $22.71 \pm 0.00$ & $23.16 \pm 0.02$ & $23.77 \pm 0.04$ & $23.07 \pm 0.07$ & $22.60 \pm 0.08$ & 1.4792\\ 
159 & $110.8490125$ & $-73.4603960$ & $21.01 \pm 0.00$ & $21.92 \pm 0.01$ & $22.42 \pm 0.01$ & $22.76 \pm 0.05$ & $22.85 \pm 0.09$ & 0.397\\ 
163 & $110.8535771$ & $-73.4596357$ & $21.55 \pm 0.00$ & $21.75 \pm 0.01$ & $22.24 \pm 0.01$ & $19.96 \pm 0.00$ & $20.02 \pm 0.01$ & 1.0815\\ 
164 & $110.8537666$ & $-73.4592641$ & $20.97 \pm 0.00$ & $21.62 \pm 0.00$ & $22.29 \pm 0.01$ & $21.95 \pm 0.02$ & $22.12 \pm 0.05$ & 1.0823\\ 
 \hline
\end{tabular}
\label{table: photometry}
\caption{Measured photometry (in AB magnitudes) of the sources plotted in Figures \ref{fig: smacs 444}, \ref{fig: smacs 770}, and \ref{fig: smacs 1000}, as well as their MUSE or NIRSpec spectroscopic redshifts from \cite{caminha+2022} or \cite{2022arXiv220708778C}, respectively. The photometry is measured in $1\arcsec$ apertures (see Section \ref{sec: smacs}). The full version of this table containing 172 objects, as well as the F356W photometry, and IDs in \cite{caminha+2022} or \cite{2022arXiv220708778C} catalogs (where available) is available online in machine-readable format.}
\end{table*}

\end{document}